\documentclass[a4paper,fleqn,usenatbib]{mnras}

\usepackage[T1]{fontenc}
\usepackage{ae,aecompl}
\usepackage{xcolor}

\usepackage{graphicx}	
\usepackage{amssymb}	
\usepackage{pdflscape}	
\usepackage[english]{babel}
\pdfoutput=1

\title[Fragmentation in magnetised self-gravitating discs]{On the fragmentation boundary in magnetised self-gravitating discs}
\author[Forgan, Price and Bonnell]{Duncan Forgan $^{1}$\thanks{E-mail:dhf3@st-andrews.ac.uk}, Daniel J. Price$^{2}$ and Ian Bonnell$^{1}$ \\
$^{1}$Scottish Universities Physics Alliance (SUPA), School of Physics and Astronomy, University of St Andrews, St Andrews KY16 9SS \\
$^{2}$Monash Centre for Astrophysics (MoCA) and School of Physics and Astronomy, Monash University, Vic. 3800, Australia 
}

\begin{document}

\date{Accepted}

\pagerange{\pageref{firstpage}--\pageref{lastpage}} \pubyear{2016}

\maketitle

\label{firstpage}

\begin{abstract}

We investigate the role of magnetic fields in the fragmentation of self-gravitating discs using 3D global ideal magnetohydrodynamic simulations performed with the \textsc{phantom} smoothed particle hydrodynamics code.

For initially toroidal fields, we find two regimes. In the first, where the cooling time is greater than five times the dynamical time, magnetic fields reduce spiral density wave amplitudes, which in turn suppresses fragmentation.  This is the case even if the magnetic pressure is only a tenth of the thermal pressure.  The second regime occurs when the cooling time is sufficiently short that magnetic fields cannot halt fragmentation.

We find that magnetised discs produce more massive fragments, due to both the additional pressure exerted by the magnetic field, and the additional angular momentum transport induced by Maxwell stresses.  The fragments are confined to a narrower range of initial semimajor axes than those in unmagnetised discs.  The orbital eccentricity and inclination distributions of unmagnetised and magnetised disc fragments are similar.   Our results suggest the fragmentation boundary could be at cooling times a factor of two lower than predicted by purely hydrodynamical models.

\end{abstract}

\begin{keywords}
accretion: accretion discs --- stars: formation --- quasars: supermassive black holes --- planets and satellites: formation -- magnetohydrodynamics (MHD)
\end{keywords}

\section{Introduction}

\noindent Self-gravitating accretion discs are expected to exist both around young stars and active galactic nuclei (AGN) (see e.g \citealt{Kratter2016} and \citealt{Rice2016} for recent reviews).  In star formation, the angular momentum distribution of dense molecular cloud cores \citep{Goodman1993} ensures that upon collapse into star-disc systems, the discs have an initial mass comparable to the protostar \citep{Lin1990,Bate2010,Tsukamoto2014a}, ensuring that self-gravity plays a role in the disc evolution.  Indeed, much of the star's final mass is expected to be drawn from the disc through efficient, globally determined angular momentum transport \citep{Laughlin1996,Kratter2010a,Forgan2011}.  In the case of AGN, the disc to central object mass ratio can be quite low (of order $10^{-3}$) but may still be self-gravitating if the disc is thin \citep{Goodman2003,Lodato2008}.

Accretion discs are prone to the gravitational instability if \citep{Toomre_1964}
\begin{equation}
Q = \frac{c_s \kappa}{\pi G \Sigma} \leq 1, \label{eq:Q}
\end{equation}
where $c_s$ is the sound speed of the gas, $\Sigma$ is the gas surface density. The quantity $\kappa$ is the epicyclic frequency,   which is equal to the angular frequency $\Omega$ if the disc is Keplerian. The `Toomre $Q$' criterion (Eq.~\ref{eq:Q}) refers to axisymmetric perturbations --- non-axisymmetric perturbations are gravitationally unstable for $Q<1.5$--$1.7$ \citep{Durisen_review}.  Unstable non-axisymmetric perturbations develop into spiral density waves thanks to the differential rotation of the disc \citep{Goldreich1965}.  These waves are trans-sonic, with Mach numbers of order unity \citep{Cossins2008}.

Depending on the available cooling mechanisms, the disc can become marginally stable \citep{Paczynski1978}.  This is a self-regulated quasi-steady state where $Q$ is maintained near the stability limit, and the local weak shock heating through spiral density waves (increasing $Q$) approximately balances the local radiative cooling (reducing $Q$).

This quasi-steady state can be expressed in terms of the dimensionless cooling parameter $\beta_c$, defined according to
\begin{equation}
\beta_c = t_{cool} \Omega,
\end{equation}
where $t_{cool}$ is the cooling time of the gas, and if the angular momentum transport is locally determined, by defining the dimensionless stress parameter $\alpha$ in terms of a pseudo-viscosity \citep{Shakura_Sunyaev_73,Gammie}:

\begin{equation}
\nu = \alpha c_s H,
\end{equation}

where $H$ is the disc scale height.  This approximates the heating of the disc as due to a turbulent pseudo-viscosity.  This is an acceptable approximation if the spiral density wave modes couple and decay into gravito-turbulence and $Q\sim 1$  \citep{Balbus1999,Gammie}.  When the disc is marginally stable and in thermal equilibrium, there is a direct relationship between this viscous stress heating and the radiative cooling, and thus
\begin{equation}
\alpha = \left|\frac{d \ln \Omega}{d \ln r}\right|^{-2}\frac{1}{\gamma(\gamma-1)\beta_c},
\end{equation}

where $\gamma$ is the ratio of specific heats.  If the radiative cooling is sufficiently vigorous (or the accretion of material into the disc is sufficiently vigorous), then this marginally stable state cannot be maintained.  This has been interpreted as a maximum stress that the gravitational instability can support, with $\alpha_{\rm max} \approx 0.06$ \citep{Rice_et_al_05}.  The gravitational stress saturates at this maximum value, and if density perturbations can continue to grow (either via strong cooling or rapid mass loading), then these perturbations become unstable to gravitational collapse, resulting in disc fragmentation.

There remains debate as to the precise value of this maximum \citep{Meru2011,Meru2012,Michael2012,Rice2012,Rice2014}, and whether fragmentation can occur under low stresses, due to stochastic density fluctuations well above the RMS value that can occur given sufficient time \citep{Paardekooper2012,Young2015,Young2016}.  It remains the case that fragmentation is \emph{favoured} in discs that can no longer sustain thermal equilibrium through heating via spiral density waves and gravito-turbulence.  Equivalently, disc fragmentation is favoured when the local Jeans mass inside a spiral density wave is decreasing rapidly \citep{Forgan2011a}, either through decreases in the local sound speed, or increases in the local gas density in the case of irradiated discs \citep{Kratter2011}.



Magnetic fields have been absent from most studies of disc fragmentation to date.  The usual argument for protostellar discs is that the low ionisation fractions mean that magnetic fields can be ignored.  This is in stark contrast to studies of protostellar disc formation, which show that magnetic fields play a crucial role in extracting angular momentum from the system via magnetic braking.  In ideal MHD simulations, this braking is so efficient that it can suppress disc formation entirely \citep{Allen2003,Price2007,Hennebelle2009a,Commercon2010}. Observations indicate that discs of order 100 au can exist around Class 0 protostars \citep[e.g.][]{Tobin2015}, and the first direct observations of disc fragmentation proceed in this regime \citep{Tobin2016}.  This apparent discrepancy between efficient magnetic braking and observed extended discs is resolved by either invoking turbulence, non-ideal MHD effects, or both \citep{Seifried2013,Tomida2015,Wurster2016}.  Other surveys have shown that massive extended discs are perhaps less common \citep{Maury2010}, suggesting that some limited magnetic braking is occurring, and consequently limiting the frequency of disc fragmentation.

Discs around AGN are also likely to possess their own fields, through intense irradiation from the central engine and the surrounding environs.  The presence of fields in both types of disc is evidenced by observations of jets and outflows, which are thought to be both launched and collimated by tightly wound field configurations \citep{Pudritz2012, Marti-Vidal2015,Rodriguez-Kamenetzky2016}.
  
There are few studies to date which consider the influence of magnetic fields on self-gravitating discs.  During the same period that \citet{Gammie} established equilibrium relationships for unmagnetised gravito-turbulence, \citet{Kim2001} investigated magnetised self-gravitating gas in 2D shearing box simulations, determining that the magnetic Toomre $Q$
\begin{equation}
Q_m = \frac{\sqrt{c^2_s + v^2_A}\Omega}{\pi G \Sigma} < 1.2-1.4,
\end{equation}
for non-axisymmetric perturbations to grow in the presence of a magnetic field (see also \citealt{Kim2003}).  Linear stability analysis of magnetised self-gravitating discs \citep{Lin2014a} shows that the magneto-rotational instability (MRI) can be either stabilised or enhanced by gravitational instability (GI).  If the resistivity is uniform, MRI modes with small radial lengths are stabilised.  If the resistivity is a function of altitude from the midplane, MRI modes with radial length of order the disc thickness can be enhanced, which can significantly enhance axisymmetric density perturbations if the field is strong and toroidal.  Interestingly, unstable modes can transition between MRI and GI, when the perturbations in gravitational potential and magnetic energy are comparable.

Fromang and collaborators \citep{Fromang2004,Fromang2004a,Fromang2005} performed 3D global simulations (assuming an isothermal disc), which suggested that weak magnetic fields (with a plasma parameter $\beta_p\sim 10^3$) would not prevent fragmentation. The growth of MRI induces turbulence, creating spiral modes that are out of phase with those produced by GI, modulating the gravitational stress and subsequently the star's accretion rate.  We will address here whether these features are due to the isothermal equation of state, or a generic feature of MRI-GI interaction in global discs.

More recently, local shearing box simulations were used to investigate the nature of 3D gravito-turbulence under the effect of weak ($\beta_p>>1$) and strong fields ($\beta_p=1$, \citealt{Riols2016}).  They identify three regimes, delineated by the energy balance between kinetic, gravitational and magnetic components.  In the regime where magnetic energy is sub-dominant, a gravito-turbulent state can still be maintained, even with a relatively high Toomre-$Q$.

In the second regime, the magnetic energy exceeds the local gravitational energy (but not the kinetic energy).  A gravito-turbulent state is still maintained, but the stresses in the system are now produced by magnetic fields, which use self-gravity as a dynamo.  The weak turbulence maintained by gravitational instability is amplified by the formation of elongated current sheets, being dissipated as heat through magnetic reconnection.  Interestingly, the deformation of the field results in ``plasmoids'', magnetic islands which are remarkably similar to disc fragments, but created by fundamentally different physics.  In the final regime, gravitational instability is completely suppressed by magnetic tension.  \citet{Riols2016} also studied the fragmentation boundary in the presence of magnetic fields.  They find for a range of magnetic field strengths, the disc can still fragment if $\beta_c <5$.

To date, there is no corresponding study of magnetic fields on disc fragmentation using 3D global simulations at sufficient resolution. To this end, we conduct a series of numerical experiments on magnetised self-gravitating accretion discs using smoothed particle hydrodynamics (SPH).  We adopt a simple parametrisation to initialise both the local cooling rate and the initial magnetic field configuration, which we describe along with the code and our analysis methods in \S \ref{sec:method}.  In \S \ref{sec:results} we describe how magnetic fields alter the structure of self-gravitating discs and the properties of fragments that they produce.  In \S \ref{sec:discussion} we discuss the implications of our results, and in \S \ref{sec:conclusions} we summarise the work.

\section{Method} \label{sec:method}



\subsection{Phantom}

\noindent Smoothed Particle Hydrodynamics (SPH) is a Lagrangian method for solving the equations of fluid dynamics.  The fluid is decomposed into a collection of particles, each possessing a mass $m_i$, position $\mathbf{r}_i$, velocity $\mathbf{v}_i$, internal energy $u_i$ and smoothing length $h_i$ (where $i$ is the particle label). The density of the fluid at any position is reconstructed using the kernel weighted estimator
\begin{equation}
\rho (\mathbf{r}) = \sum^{N}_{i=1} m_i W(\left|\mathbf{r} - \mathbf{r}_i\right|, h),
\end{equation}
where $W$ is the smoothing kernel.  The kernel function is selected to have compact support within the range  $\left|\mathbf{r} - \mathbf{r}_i\right| = [0,2h]$, so that $N$ represents the number of neighbouring particles within a distance $2h$ of $\mathbf{r}$.

The equations of motion for the fluid proceed entirely from this density estimator (with an appropriate Lagrangian and variational principle), yielding a consistent framework for solving the (magneto)hydrodynamic equations (see \citealt{Price2012} for a review).  We use the SPMHD code \textsc{Phantom}. The implementation of MHD in \textsc{Phantom} follows the basic scheme described in \citet{Price2004,Price2004a,Price2005a} (see review by \citealt{Price2012}) with the divergence constraint on the magnetic field enforced using the constrained hyperbolic divergence cleaning algorithm described by \citet{Tricco2012} and \citet{Tricco2016}. We assume ideal MHD for this work ---  the code is capable of non-ideal MHD (see e.g. \citealt{Wurster2016}) but this is beyond the scope of the present paper.

 We employ artificial viscosity, conductivity and resistivity to resolve shocks and prevent unphysical particle interpenetration, for viscosity adopting the time-dependent viscosity of \citet{Morris1997}, where the $\alpha_{SPH}$ can vary between 0.1 and 1, and the corresponding non-linear viscosity term is fixed at $\beta_{SPH}=2$. The particles evolve on individual timesteps, and the gravity forces are computed using a binary tree similar to that described in \citet{Gafton2011}.

The central object is represented by a sink particle \citep{Bate_code}, which accretes SPH particles which stray within a distance $r_{accrete}$, provided the SPH particle is bound and possesses an angular momentum less than that required to orbit beyond $r_{accrete}$.  Particles that stray within $r_{accrete}/2$ are accreted without any checks.  Dynamic sink creation is turned off, as our simulations can easily resolve the individual fragments (see section \ref{sec:ICs}).

\subsection{Calculating Disc Stresses \label{sec:stresses}}

\noindent To compare the relative strength of self-gravity and magnetic fields in the steady state, we compute their associated stresses via the dimensionless $\alpha$-parameter.  This assumes that the angular momentum transport is locally determined, which is appropriate for self-gravitating discs when the disc to central object mass ratio is sufficiently low \citep{Forgan2011}.  For each process, we compute the viscous stress tensor $T_{r\phi}$, which is related to $\alpha$ according to:

\begin{equation}
T_{r\phi} = \left|\frac{d \ln \Omega}{d \ln r}\right| \alpha \Sigma c^2_s,
\end{equation}

\noindent i.e. $T$ is proportional to the local pressure.  Once $T$ is computed, we simply invert for $\alpha$ using the vertically averaged sound speed and $\Sigma$.  The viscous stress produced by gravito-turbulence can be written \citep{Lynden-Bell1972}
\begin{equation}
T_{r\phi} = \int \frac{g_r g_\phi}{4\pi G} dz,
\end{equation}
where $g_r$ and $g_{\phi}$ are the radial and tangential components of the gravitational acceleration respectively.  The  Maxwell stress induced by the magnetic field is
\begin{equation}
T_{r\phi} = \int \frac{B_r B_\phi}{\mu_0} dz,
\end{equation}
where again $B_r$ and $B_\phi$ are radial and tangential field components.  Finally, a third contribution to the total stress arises from the velocity and density perturbations produced by both the magnetic and gravitational fields - the Reynolds stress
\begin{equation}
T_{r\phi} = \Sigma \delta v_r \delta v_\phi,
\end{equation}
where each $\delta v$ represents deviations from the mean flow: e.g. $\delta v_r = v_r - \left<v_r \right>$.

Artificial viscosity also induces angular momentum transport, and as such we should calculate it to confirm that our solutions are not dominated by numerics.  This effective viscosity can be represented as \citep{Artymowicz1994,Murray1996,Lodato2010}
\begin{equation}
\nu_{\rm art} = \alpha_{\rm art} c_s H = \frac{1}{10} \alpha_{\rm AV} c_s h,
\end{equation}
relating the user-defined artificial viscosity parameter $\alpha_{AV}$ to the resulting stress, represented by $\alpha_{art}$
\begin{equation}
\alpha_{\rm art} = \frac{1}{10} \alpha_{\rm AV} \frac{<h>}{H},
\end{equation}
where $<h>$ is the azimuthally averaged smoothing length. As might be expected, discs with poorly resolved vertical structure will be dominated by artificial viscosity.  As $H$ decreases with decreasing $r$, the inner regions of our discs ($r \lesssim 7$) will be dominated by numerical viscosity.

\subsection{CLUMPFIND analysis of Disc Fragments}

We wish to identify how the properties of the fragments themselves change as the magnetic field is varied.  We apply the CLUMPFIND algorithm to detect local minima in the gravitational potential \citep{Smith2009}.  This algorithm places each particle into a clump depending on its gravitational potential $\phi$, and that of the particles in its neighbour sphere.

The procedure is straightforward:  All particles are sorted into a list of decreasing $\left|\phi \right|$, and are initially unassigned to any clumps.  The sink particle representing the central star begins the first clump (along with its neighbour particles).  The first particle on the list is then considered.  If one of its neighbours is in clump $j$, then the particle is added to clump $j$.  If the particle's neighbours belong to multiple clumps, then the particle is added to the clump to which the majority of its neighbours belongs.  If the particle has no neighbours in a clump, it starts a new clump and adds its neighbours to it.  We then move down the particle list to the next entry, and the process is repeated.

This procedure continues until all particles are tested.  We have several choices in how we perform this analysis.  As we wish to identify the smallest possible clumps, we do not include the contribution to the potential from the sink particle. If we had, then smaller potential wells would not be detected by the algorithm.  Secondly, we do not stipulate that our clumps must be gravitationally bound (the basic CLUMPFIND algorithm is agnostic to a clump's boundness, but can be augmented to determine the bound component by removing outer particles until the clump's total energy is negative).

This gives a set of clumps for each timestep, but it does not inform us how to track a clump over several timesteps.  To do this, we use a standard algorithm from halo tracking in cosmological simulations (cf \citealt{Springel2001a}).  For a clump $i$ in timestep 1 and a clump $j$ in timestep 2, the two clumps are shown to be the same clump if:

\begin{enumerate}
\item The most bound particle in clump $i$ is also in clump $j$, and
\item At least 50\% of the particles in clump $i$ are also in clump $j$
\end{enumerate}

We can then use this clump tracking system to measure mass evolution of the fragment, as well as orbital data.

\section{Results} \label{sec:results}

\subsection{Initial Conditions}\label{sec:ICs}

\noindent We adopt initial conditions similar to those employed by \citet{Rice_et_al_05} and many subsequent authors.  The simulations are scale-free.  The disc is composed of $N_{part}$ SPH (gas) particles, with a surface density profile $\Sigma \propto r^{-1}$, sound speed profile $c_s \propto r^{-0.75}$, extending between $r_{in}=0.25$ and $r_{out}=25$, with a total mass $M_d=0.25$.  The central object is represented by with a sink particle, mass $M=1$, placed initially at the origin.  The sink is free to move, with $r_{accrete}=0.25$.

We assume an adiabatic equation of state, with the ratio of specific heats $\gamma=5/3$. We adopt the usual $\beta$-cooling formalism
\begin{equation}
\dot{u}_{\rm cool} = -\frac{u}{t_{\rm cool}} = -\frac{u\Omega}{\beta_c},
\end{equation}
where we fix $\beta_c={\rm const}.$ throughout the disc.  We initialise a toroidal magnetic field with strength given by a  plasma parameter $\beta_p$ which is initially constant with radius
\begin{equation}
\beta_p = \frac{2P\mu_0}{B^2},
\end{equation}
i.e., the ratio of thermal to magnetic pressure.  The magnetic field is allowed to evolve freely under ideal MHD conditions, with no external field applied. 

To resolve fragmentation correctly, we must ensure that the Jeans mass is larger than a neighbour-group of particles \citep{Burkert_Jeans}, i.e.
\begin{equation}
M_{J,{\rm resolve}} = 2M_{d} \frac{N_{\rm neigh}}{N_{\rm part}}.
\end{equation}
For our simulations, the minimum Jeans mass resolvable is $M_{J,{\rm resolve}}= 5\times10^{-5}$ for $N_{\rm part}=5\times 10^5$, and $M_{J,{\rm resolve}}= 2.5\times10^{-6}$ for $N_{\rm part}=10^6$.  As we will see in the following sections, the typical fragment mass is of order $10^{-3}$, which is well above the resolution limit.

This resolution is only guaranteed to be sufficient where the disc evolution is also well resolved, i.e. where the artificial viscosity in the simulation is low compared to the pseudo-viscosity generated by the gravitational and magnetic fields, as described in section \ref{sec:stresses}.  If not, self-gravitating fragments may not form due to excessive artificial angular momentum transport.


\subsection{Steady State Discs \label{sec:steadydisc}}

\begin{figure*}
\begin{center}$\begin{array}{cc}
\includegraphics[scale=0.3]{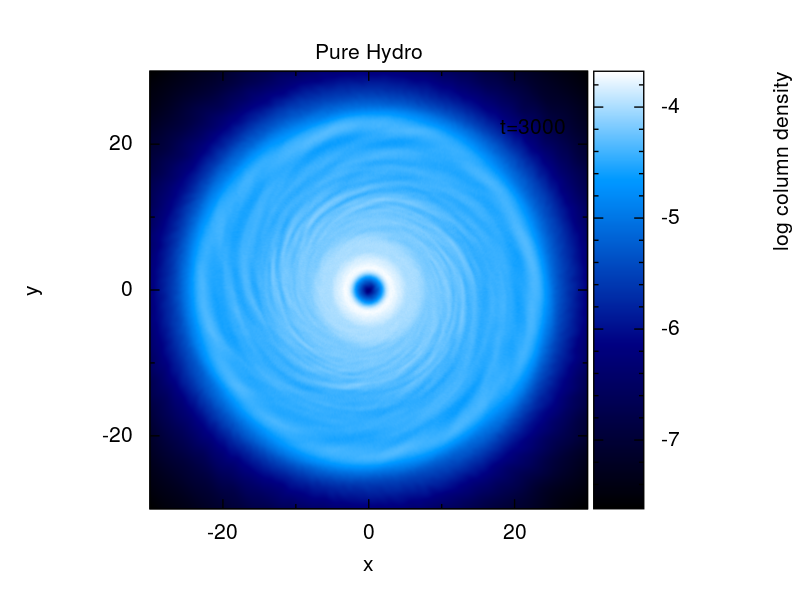} &
\includegraphics[scale=0.3]{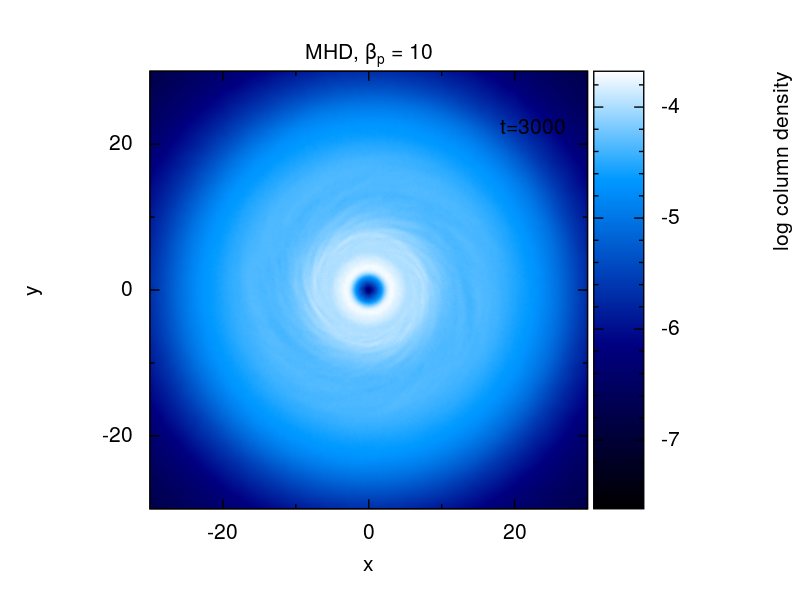} \\
\end{array}$
\end{center}
\caption{Comparison of surface density between the unmagnetised (left) and magnetised (right; with $\beta_p=10$) calculations, using $\beta_c=9$. The disc is smoother, with weaker spiral shocks in the magnetised case \label{fig:beta_9coldens}}
\end{figure*}

Before assessing fragmenting systems, we perform two calculations with $N_{\rm part}=10^6$ and $\beta_c=9$ to investigate the steady-state behaviour of non-fragmenting self-gravitating discs with and without magnetic fields. Visually inspecting both discs at the same timestep (Figure \ref{fig:beta_9coldens}), it is immediately obvious that magnetic fields attenuate the gravitational instability in the outer region.  The mean column density structure appears to be similar in both cases, but the spiral density waves present throughout the unmagnetised disc are suppressed in the magnetised disc at $r>15$.

\begin{figure*}
\begin{center}$\begin{array}{cc}
\includegraphics[scale=0.4]{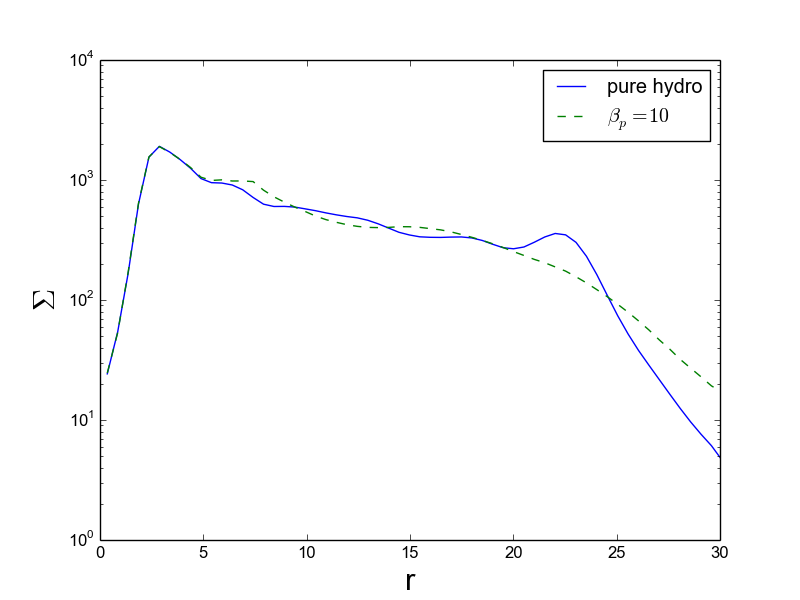} &
\includegraphics[scale=0.4]{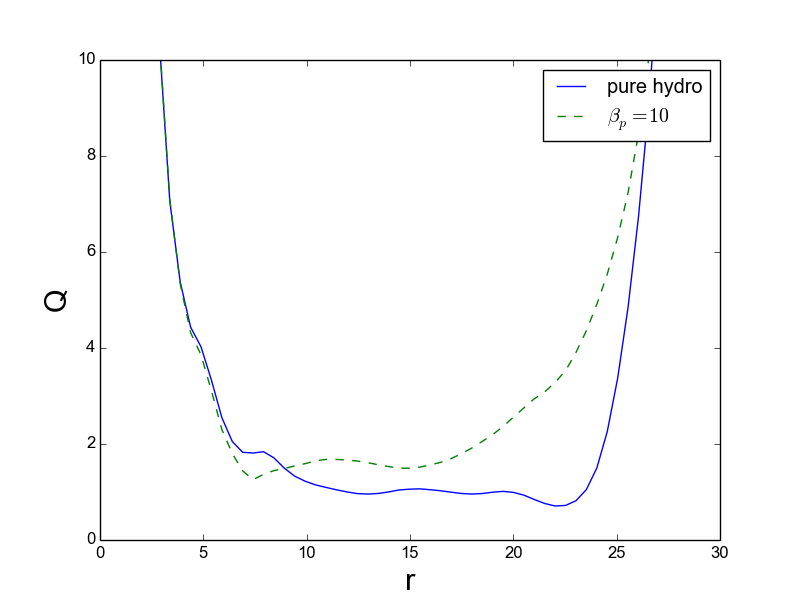} \\
\end{array}$
\end{center}
\caption{Comparison of steady state disc structures in non-fragmenting self-gravitating discs.  We show time-averaged surface densities (left) and time averaged Toomre $Q$ parameter (right) in the magnetised (dashed line) and unmagnetised (solid line) cases. \label{fig:beta_9disc}}
\end{figure*}

\begin{figure*}
\begin{center}$\begin{array}{cc}
\includegraphics[scale=0.4]{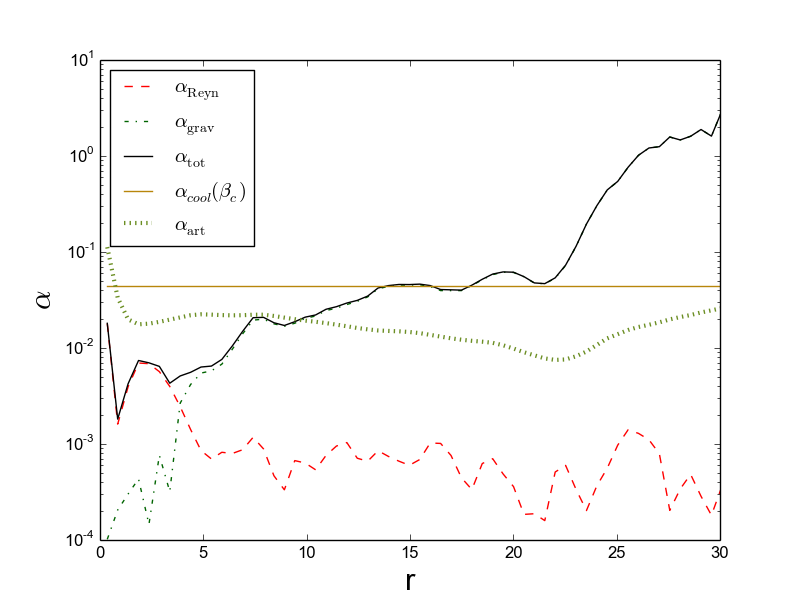} &
\includegraphics[scale=0.4]{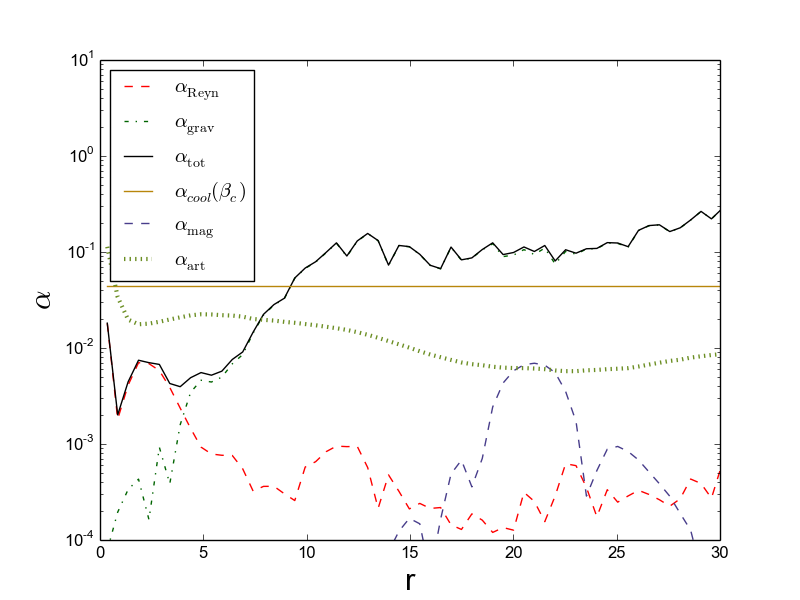} \\
\end{array}$
\end{center}
\caption{Angular momentum transport in non-fragmenting self-gravitating discs.  Left panel shows the time-averaged $\alpha$-parameters for an unmagnetised disc with $\beta_c=9$, while right panel shows the same disc with $\beta_c=9$ and $\beta_p=10$.  In both cases we also plot the predicted $\alpha(\beta_c)$ for an unmagnetised self-gravitating disc in thermal equilibrium. \label{fig:beta_9alpha}}
\end{figure*}

We can confirm this by inspecting the time-averaged surface density and Toomre $Q$ parameter (Figure \ref{fig:beta_9disc}).  Adding a toroidal field (with strength given by $\beta_p=10$) does not significantly affect the surface density profile between $r=5-20$ , but does result in a more shallow profile beyond $r=20$.  The sound speed in the disc increases noticeably beyond $r=10$ (where Maxwell stresses begin to become significant, see Figure \ref{fig:beta_9alpha}).  As the rotation curves of both discs are very similar, the resulting Toomre $Q$ for the magnetised disc increases sharply beyond $r=15$.  If one considers the magnetic Toomre $Q$ parameter
\begin{equation}
Q_m = \frac{\sqrt{c^2_s + v^2_A}\Omega}{\pi G \Sigma} = Q \sqrt{1 + \frac{1}{\beta_p}},
\end{equation}
a gravitationally unstable magnetised disc would still possess a standard Toomre $Q$ close to 1.5-2 for $\beta_p=10$.  The magnetic stresses have therefore caused sufficient heating to outmatch the local $\beta_c$-cooling, and quell the gravitational instability in this disc.

This is also reflected in the measured angular momentum transport (Figure \ref{fig:beta_9alpha}).  We plot the $\alpha$-parameter for each component, and the total
\begin{equation}
\alpha_{\rm tot} = \alpha_{\rm Reyn} + \alpha_{\rm grav} + \alpha_{\rm mag}.
\end{equation}
We also plot the predicted total $\alpha$, $\alpha_{\rm cool}$ given $\beta_c=9$ and $\gamma =5/3$ (for an unmagnetised disc)
\begin{equation}
\alpha_{\rm cool}(\beta_c) = \frac{4}{9 \gamma(\gamma-1)\beta_c} = 0.0444.
\end{equation}

The unmagnetised disc (left plot in Figure \ref{fig:beta_9alpha}) possesses an $\alpha_{\rm tot}$ close to that of the predicted value between $r\sim 12-22$, which corresponds neatly to the minima in $Q$ demonstrated in the right panel of Figure \ref{fig:beta_9disc}.  In the magnetised case, the disc exhibits a total $\alpha\sim 0.1$, quite in excess of the unmagnetised prediction.  However, we can see that this is not due to the significant increase of adding a non-zero $\alpha_{\rm mag}$, or an enhanced $\alpha_{\rm Reyn}$.  In fact, there are indications that $\alpha_{\rm Reyn}$ is decreased compared to the unmagnetised case, presumably due to weaker spiral density wave structures failing to generate the same strength of velocity perturbations.

The increased total stress must be purely from gravitational forces, and appears to be due to the modified disc structure produced by adding magnetic fields.  The magnetised disc maintains a shallower surface density profile, and hence the gravitational torques exerted by the outer disc become more significant in the inner parts.  Large scale magnetic fields are also likely to be mediating torques and boosting stresses.  These fields are also producing weak outflows, resulting in a negligible amount of mass at high altitude compared to the unmagnetised case.

The Toomre-$Q$ in these regions remains approximately 2 and larger, ensuring that the disc is stable against gravitational perturbations, even when the gravitational stresses remain high.  We should therefore view this enhanced stress as resulting from long-range torques from the outer disc, rather than locally generated gravito-turbulence.  As long-range torques can generate $\alpha$ values well in excess of the fragmentation boundary without producing fragments \citep[cf][]{Harsono2011}, the disc can remain stable under such perturbations.

\subsection{Fragmenting Discs}


\begin{figure*}
\begin{center}$\begin{array}{cc}
\includegraphics[scale=0.3]{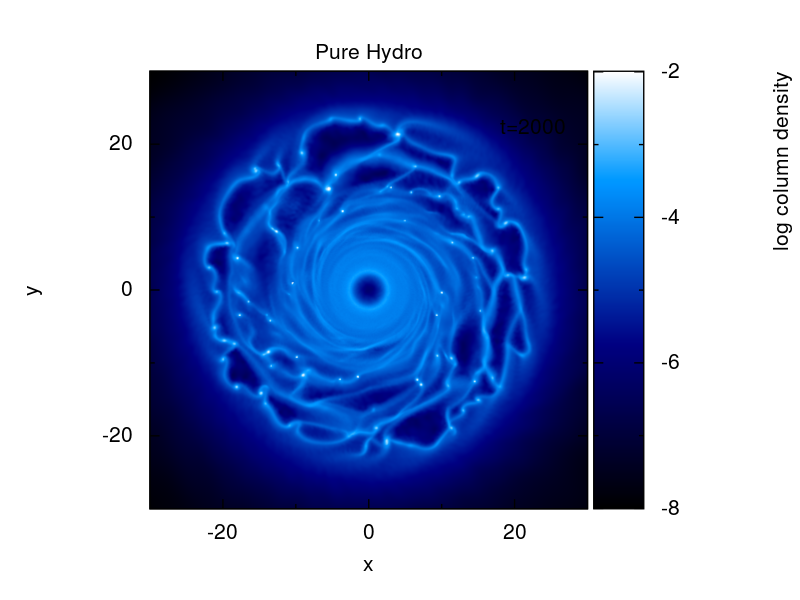} &
\includegraphics[scale=0.3]{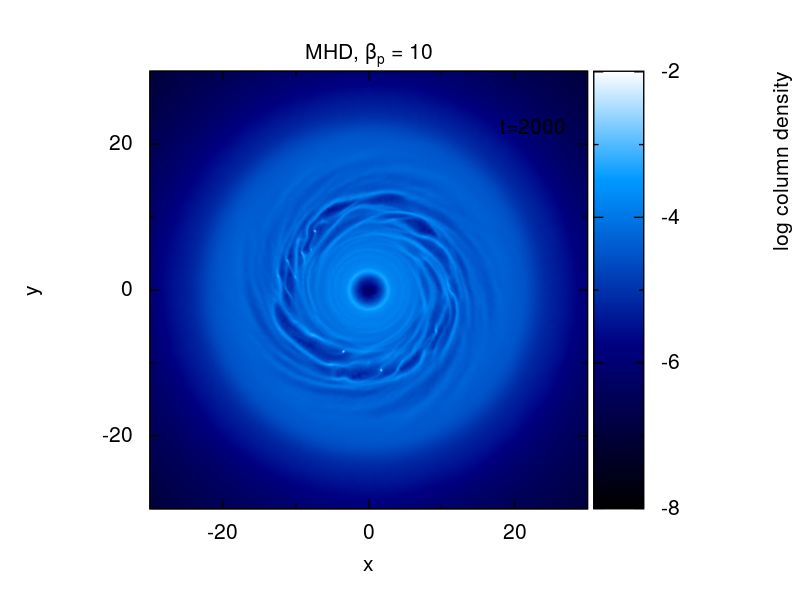} \\
\end{array}$
\end{center}
\caption{Comparison of fragmentation in unmagnetised (left) and magnetised (right; $\beta_p = 10$) discs.  Both plots have a $\beta_c=4$ and $N_{\rm part}=5\times10^5$ and are shown at $t=2000$.  \label{fig:beta_4coldens}}
\end{figure*}

\begin{figure*}
\begin{center}$\begin{array}{cc}
\includegraphics[scale=0.4]{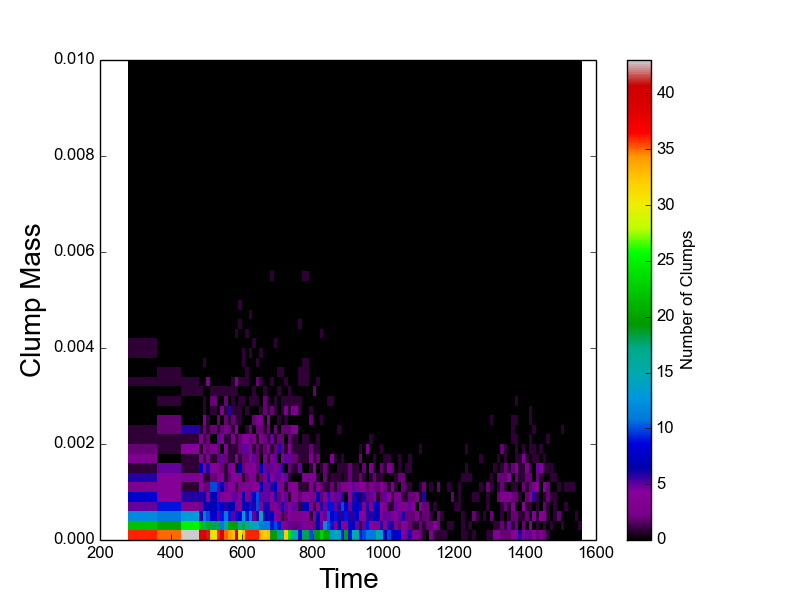} &
\includegraphics[scale=0.4]{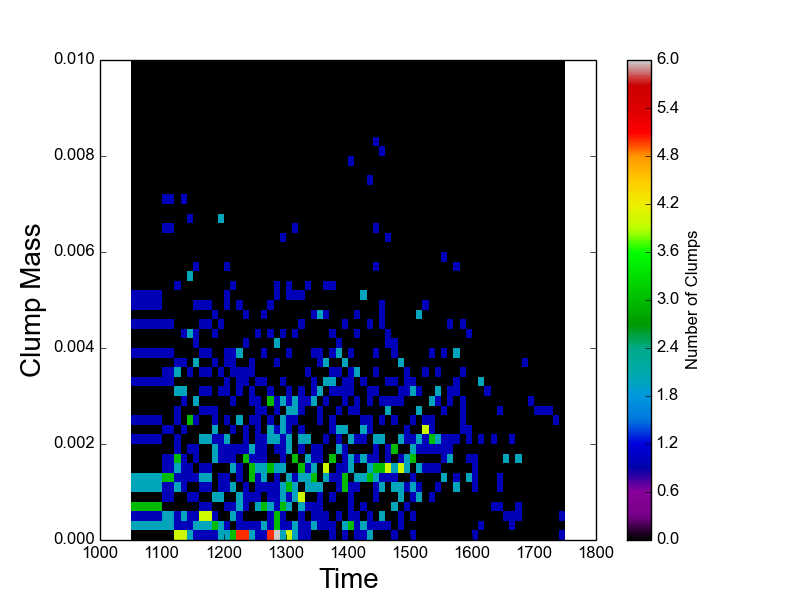} \\
\end{array}$
\end{center}
\caption{Distribution of fragment mass as a function of time, for simulations with $\beta_c=4$, $N_{part}=10^6$. Left shows the unmagnetised case, and right the $\beta_p=10$ case \label{fig:beta_4clumpmass}}
\end{figure*}

 If we now reduce the cooling time into the fragmentation regime, such that $\beta_c=4$, we might expect the outer regions of the disc to remain stable against fragmentation.  Once more, visual comparison of unmagnetised and magnetised fragmenting discs confirms this hypothesis (Figure~\ref{fig:beta_4coldens}).  Both discs have a low $\beta_c=4$.  The unmagnetised disc (left) fragments promptly, at all radii where $Q$ reaches the marginally stable regime.  The same is true for the magnetised disc, but the outer regions are no longer in the marginally stable regime, as $Q$ increases above 2 beyond $r\sim 15$, in a manner similar to the stable magnetised disc presented in section \ref{sec:steadydisc} (see Figure \ref{fig:beta_9disc}).

And what of the fragments themselves? We use the CLUMPFIND algorithm to study the properties of the fragments produced. Figure \ref{fig:beta_4clumpmass} shows the mass distribution of fragments as a function of simulation time. In the unmagnetised case, short-lived fragments initially form with masses in the range $m=0.001$--$0.004$ (if the mass is given to be in Solar units, this is approximately 1--4 Jupiter masses).  Fragments above $m=0.001$ are more persistent, surviving several orbits before migrating inwards to be tidally disrupted.  When the initial plasma $\beta_p=10$, the number of fragments produced is significantly lower.  The masses of these fragments is also somewhat larger, extending towards $m=0.006$--$0.008$.  This is due to an effective increase in the Jeans mass in the presence of the magnetic field, and the global redistribution of disc material due to Maxwell stresses (see Discussion).

\begin{figure*}
\begin{center}$\begin{array}{cc}
\includegraphics[scale=0.4]{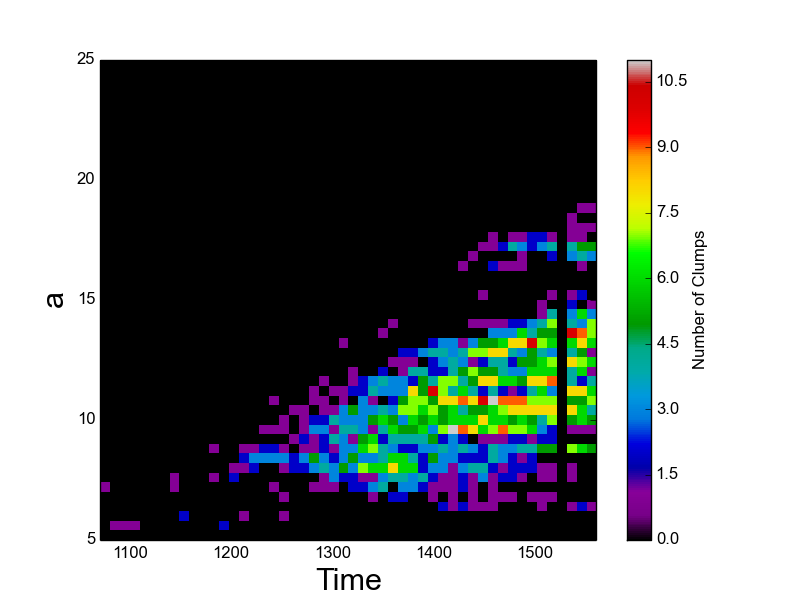} &
\includegraphics[scale=0.4]{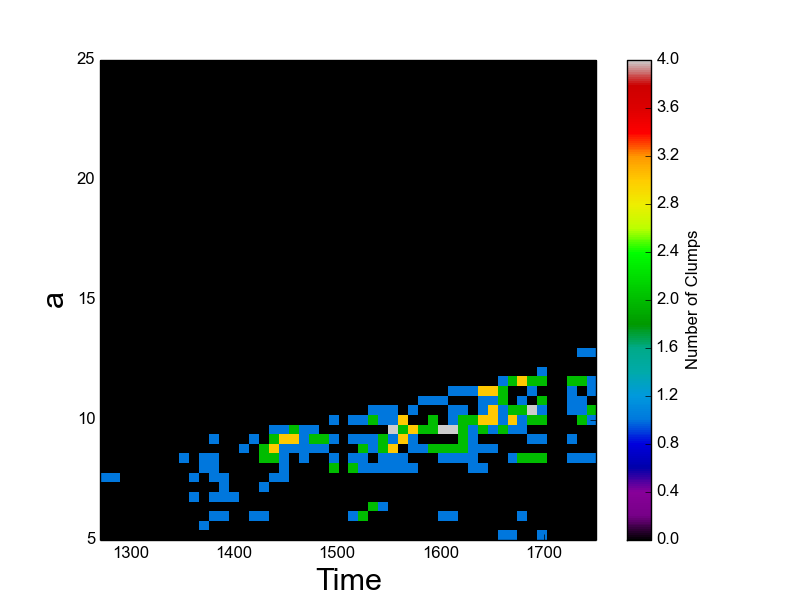} \\
\includegraphics[scale=0.4]{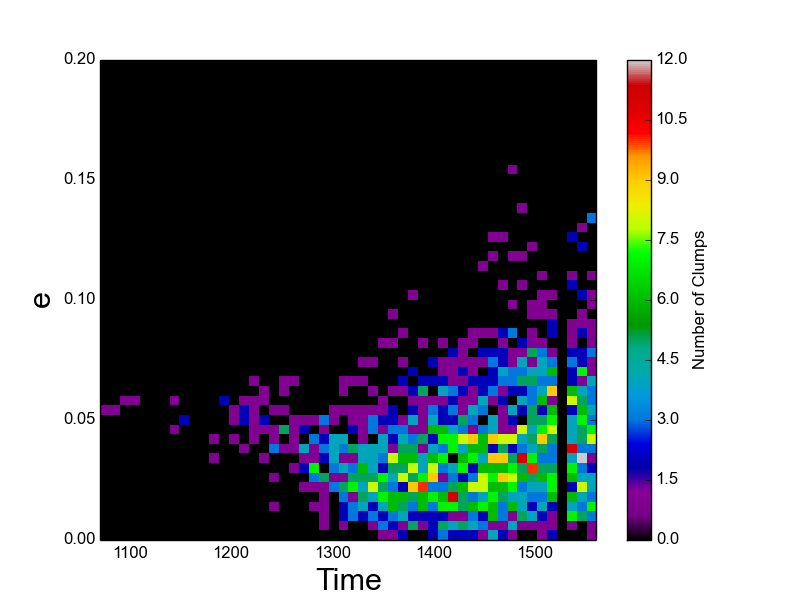} &
\includegraphics[scale=0.4]{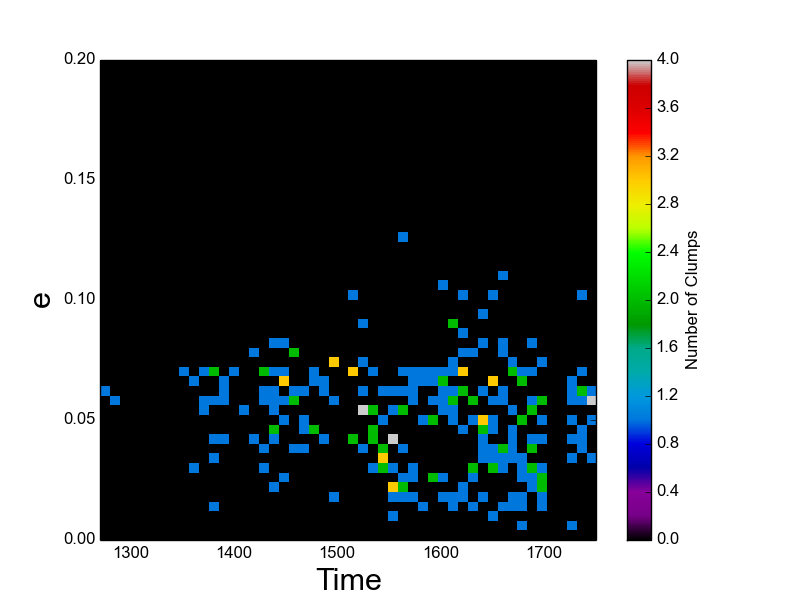} \\
\includegraphics[scale=0.4]{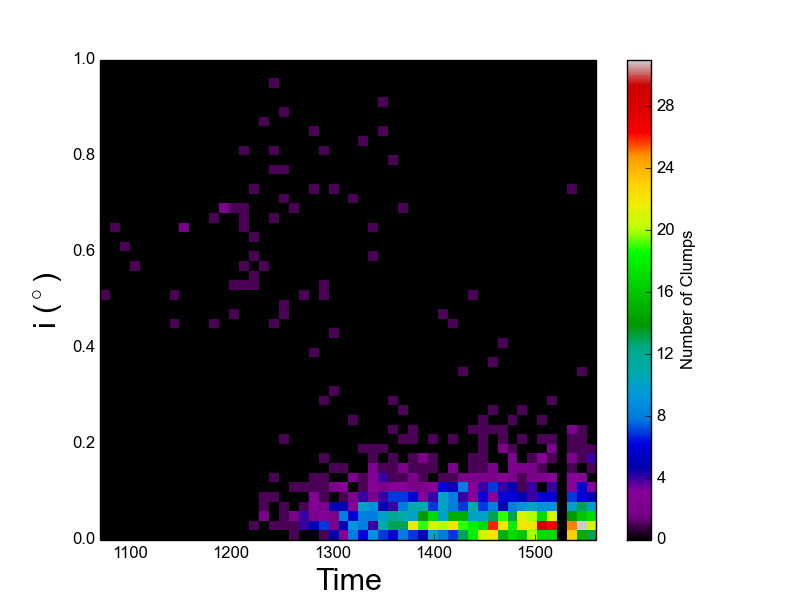} &
\includegraphics[scale=0.4]{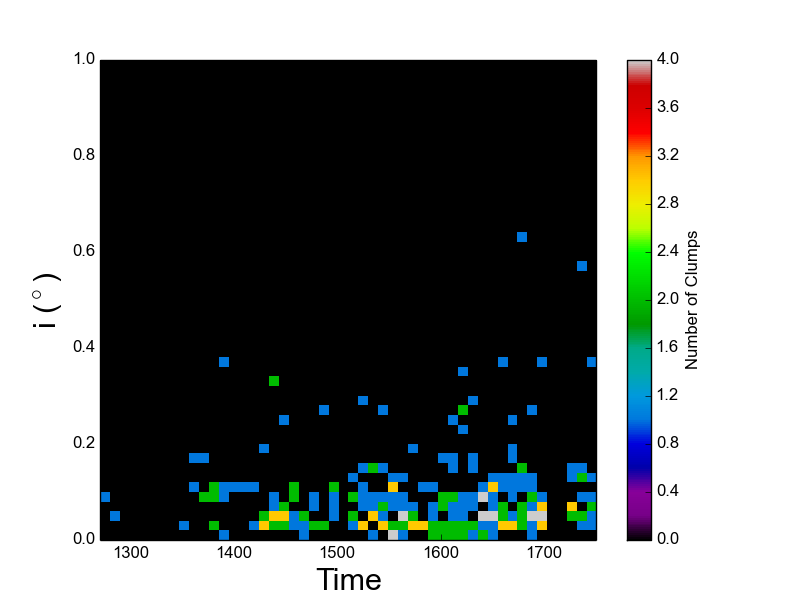} \\
\end{array}$
\end{center}
\caption{The distribution of fragment orbital elements as a function of time,for simulations with $\beta_c=4$, $N_{part}=10^6$. The left column shows the unmagnetised case, and the right the $\beta_p=10$ case. Top row: semi-major axis, middle row: eccentricity, bottom row, orbital inclination. \label{fig:beta_4orbit}}
\end{figure*}

 The semimajor axis distribution of the fragments reflects the regions of the disc that remain gravitationally unstable (top row of Figure \ref{fig:beta_4orbit}).  The unmagnetised disc is unstable from around $r=5$ to its maximum extent, and fragments promptly from around $r=10$ outwards to $r=20$.  The magnetised disc is constrained to fragment within a narrower radius limit, as the outer disc cannot maintain a sufficiently low $Q$ to permit fragmentation.

In both cases, the orbital eccentricities of the fragments are typically low (middle row of Figure \ref{fig:beta_4orbit}), and the underlying distribution is similar for both unmagnetised and magnetised discs.  Individual fragment tracking shows that the eccentricity of fragments tends to increase with time (as semimajor axis decreases).  This is presumably due to interactions with other fragments and the local disc structure extracting angular momentum and allowing inward migration towards tidal disruption.  

Most fragments form very close to the orbital plane, whether magnetic fields are active or otherwise (bottom row of Figure \ref{fig:beta_4orbit}).  Again, the inclination distribution appears to be insensitive to the magnetic field, with the majority of all fragments forming within 0.2$^\circ$ of the disc midplane.  

\subsection{Magnetic Fields as a Fragmentation Suppressor}

\begin{figure*}
\begin{center}$\begin{array}{cc}
\includegraphics[scale=0.3]{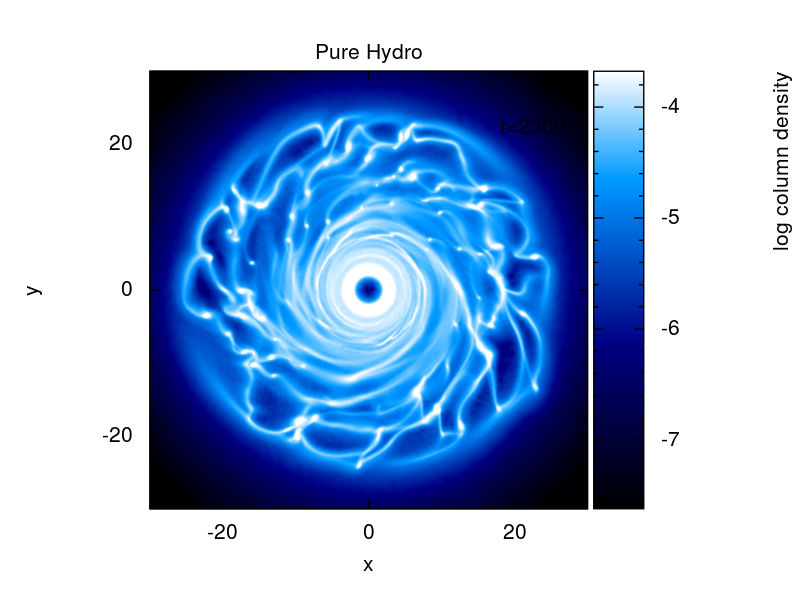} 
\includegraphics[scale=0.3]{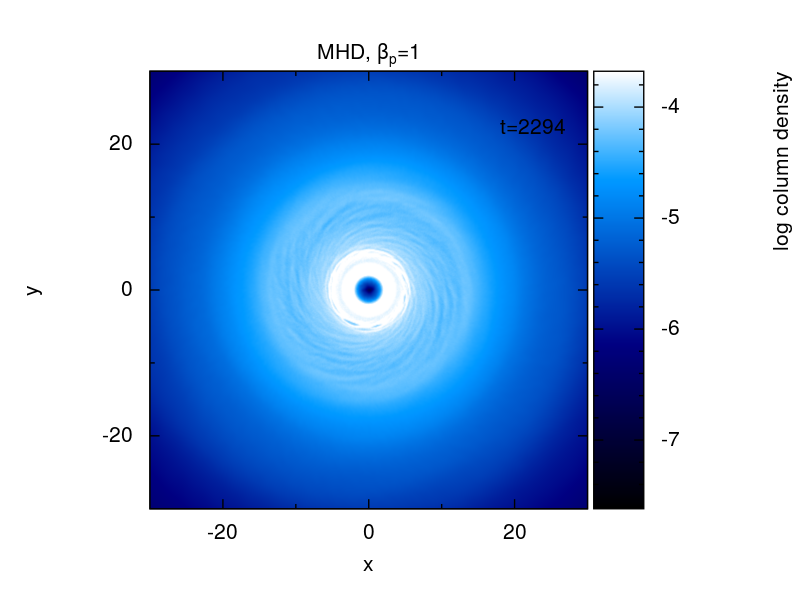}
\end{array}$
\end{center}
\caption{Suppression of fragmentation in strongly magnetised discs (comparing left and right panels), showing the calculations with $N_{\rm part}=10^6$ and $\beta_c=5$.  The left is unmagnetised, the right has a strong magnetic field ($\beta_p=1$).  While non-axisymmetric structure is evident in the form of spiral density waves, these waves have a small perturbation amplitude and do not fragment.  \label{fig:beta5_beta1_coldens}}
\end{figure*}

Imposing strong magnetic fields can suppress fragmentation if the cooling time is sufficiently long, for example if we set $\beta_c=5$, $\beta_p=1$.  Visual inspection of the surface density structure compared to the unmagnetised control run (Figure \ref{fig:beta5_beta1_coldens}) shows that spiral structures can still exist in the inner disc.  The outer disc is extremely smooth, with no signs of fragmentation.

\begin{figure*}
\begin{center}$\begin{array}{cc}
\includegraphics[scale=0.4]{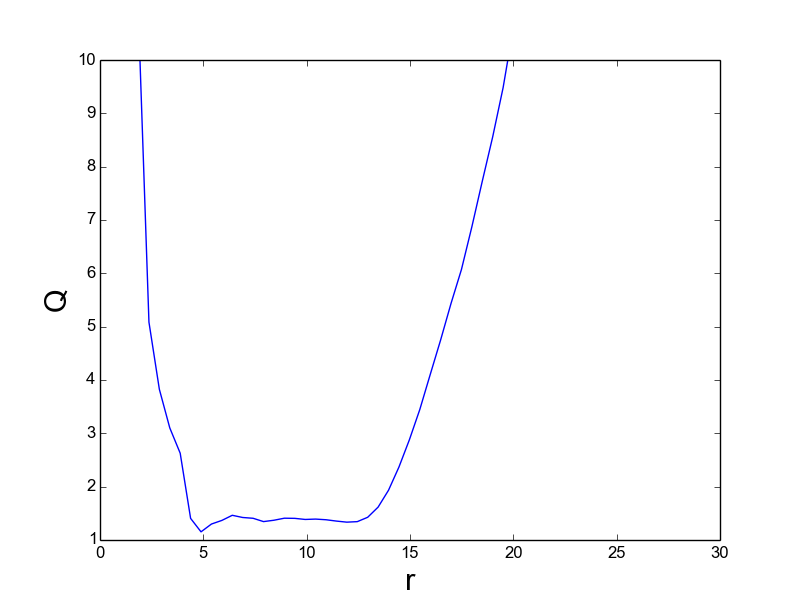} 
\includegraphics[scale=0.4]{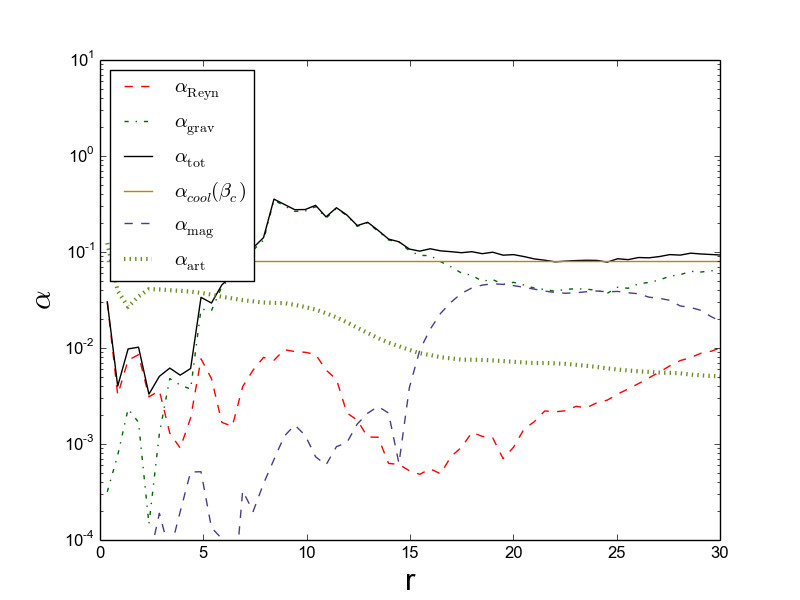}
\end{array}$
\end{center}
\caption{Toomre-$Q$ profile (left) and angular momentum transport parameters (right) in the strongly magnetised case ($\beta_c=5$, $\beta_p=1$).  The outer regions of the disc are too hot to be gravitationally unstable.  The disc maintains local thermodynamic equilibrium by equipartion of gravitational and magnetic stresses. \label{fig:beta5_beta1_Q_alpha}.}
\end{figure*}

The Toomre-$Q$ parameter confirms that only the inner radii ($r=5-15$) can be considered to be gravitationally unstable (left panel of Figure \ref{fig:beta5_beta1_Q_alpha}).  The gravitational stress dominates angular momentum transport, and far exceeds the value predicted by local equilibrium.  Despite this, the disc does not fragment.  The magnetic field does not appear to be strong enough in this regime to prevent fragmentation, but it may be the case that the stable outer disc is generating strong torques in the inner disc, invalidating the local transport approximation.

If we now look beyond $r=15$, the Maxwell stress increases, causing significant heating (right panel of Figure \ref{fig:beta5_beta1_Q_alpha}).  Interestingly, the disc maintains a total $\alpha$ that is close to the value predicted by local thermodynamic equilibrium for our imposed $\beta_c$, suggesting that angular momentum transport is still being mediated by turbulent pseudo-viscosity, where the turbulence is generated by (roughly equal) gravitational and magnetic stresses.  The gravitational stress in this regime is $\sim 0.05-0.06$ - in the absence of a magnetic field, the total $\alpha$ would lie just below the critical value for fragmentation.  This indicates that magnetic fields suppress disc fragmentation by reducing gravitational stresses, as found by both local and global simulations \citep{Fromang2005,Riols2016}.

\subsection{The Fragmentation Boundary for Magnetised Discs}

\begin{figure*}
\begin{center}$\begin{array}{cc}
\includegraphics[scale=0.4]{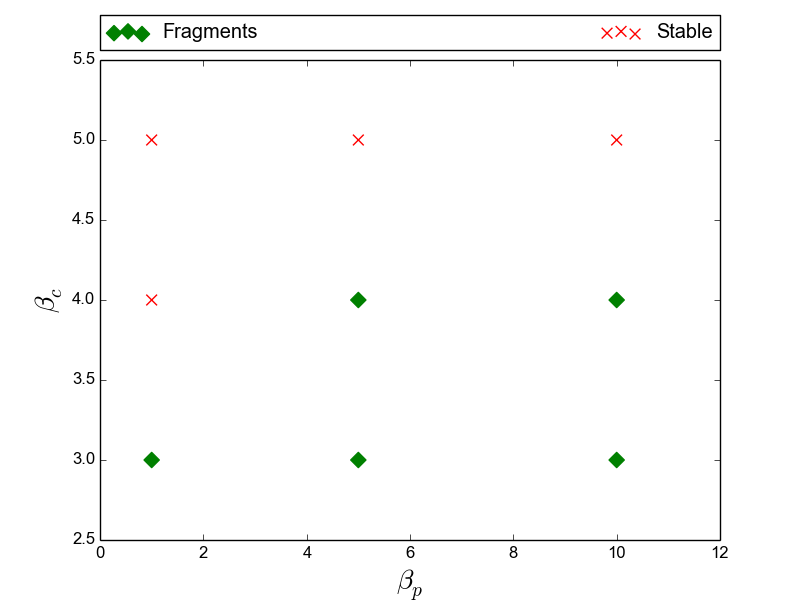} &
\includegraphics[scale=0.4]{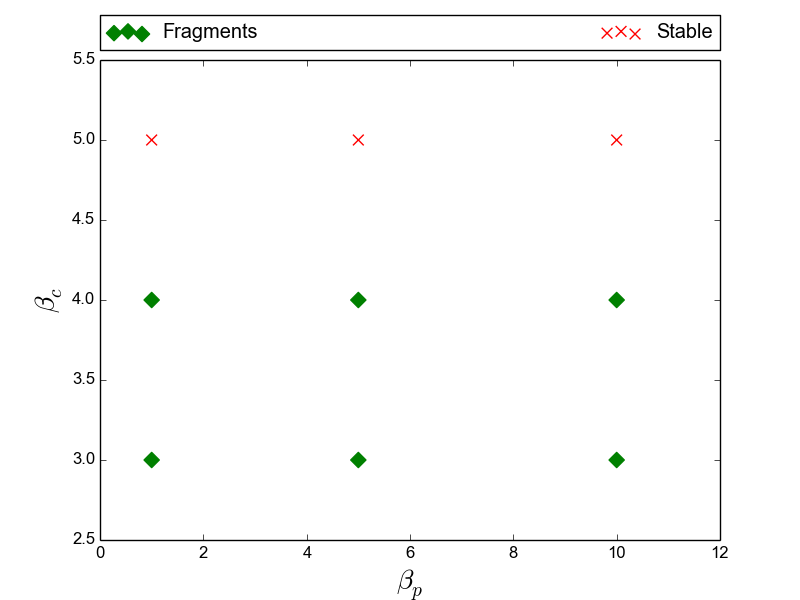} \\
\end{array}$
\end{center}
\caption{The fragmentation boundary in magnetised self-gravitating discs.  Left panels shows results from calculations with $N_{\rm part}=5\times 10^5$, while right panels shows results wit\rm h $N_{part}=10^6$.  Note the change in status for runs with $\beta_c=4$, $\beta_p=1$.  All of the above simulations fragment in the absence of magnetic fields.\label{fig:beta_vs_beta}}
\end{figure*}

\noindent Finally, we investigate the propensity of discs to fragment as a function of $(\beta_c,\beta_p)$.  Figure \ref{fig:beta_vs_beta} shows the resulting behaviour of discs in this parameter space, at two different resolutions: $N_{part}=5\times10^5, 10^6$.  Note that without magnetic fields, all of the discs studied fragment promptly.

Addition of magnetic fields at $\beta_c=5$ completely suppresses fragmentation, even at $\beta_p=10$. This is in accordance with the results of the previous section, which indicate that magnetic fields reduce the gravitational stress in the outer disc below the limit required for prompt fragmentation, and increase $Q$.  Given that in the strongest field scenario, the gravitational stress for a $\beta_c=5$ disc is about 0.05 (Figure \ref{fig:beta5_beta1_Q_alpha}), with the prompt fragmentation boundary for unmagnetised discs lying close to $\alpha = 0.06$ \citep{Rice_et_al_05}, it seems reasonable that a slight decrease in $\beta_c$ from 5 to 4 would be sufficient to raise the gravitational stress so that the total $\alpha$ exceeds the fragmentation limit.

\section{Discussion} \label{sec:discussion}

 Our principal result --- that magnetic fields can suppress fragmentation for cooling times above five times the dynamical time--- is in broad agreement with the local simulations conducted by \citet{Riols2016}.   They also show fragmentation is suppressed for $\beta_p=8$ if $\beta_c>5$.  We should however note that global and local simulations must be compared carefully.  Our global simulations show fragmentation suppression due not only to direct action of the magnetic field, but also to redistribution of the disc material diluting the gravitational instability.  Local simulations typically strictly conserve mass within the domain of interest, in effect preventing global redistribution of mass and energy as we observe. 

What values of $\beta_p$ might we expect in real discs? \citet{Wurster2016} report values between 10-1000, from the inner and outer disc respectively.  This is true both for their ideal and non-ideal MHD simulations of protostellar disc formation, and appears to be confirmed by other work \citep[e.g.][]{Tomida2015}.  This would suggest that disc fragmentation is not prevented by magnetic fields, provided the disc cooling rate is sufficiently strong locally.  If the system retains enough angular momentum to form an extended disc, this seems quite possible \citep{collapses}.

This suppression of fragmentation at what is quite low $\beta_c$, by typical values of $\beta_p$, is important to considerations of convergence in unmagnetised simulations.  If the boundary for prompt fragmentation is $\beta_c\sim 7-8$ for zero field, and potentially higher for stochastic fragmentation, then the addition of what is quite a weak field reduces the critical $\beta_c$ significantly, and requires quite strong cooling for fragmentation to occur at all.  This suggests that discussions of whether the prompt fragmentation boundary truly is $\beta_c=7,8$ or even 9 might not be particularly relevant to real self-gravitating discs, even if their magnetic fields are weak. 

Our simulations indicate that there is both a minimum and maximum radius for fragmentation to occur when magnetic fields are active, but we should note that this could be an artifact of our idealised cooling.  As $\beta_c$ declines with radius in realistic protostellar discs, we should therefore expect fragmentation to be further restricted to regions where $\beta_c$ is very low and $Q$ is maintained at the marginal stability limit, i.e. radii beyond 30 au.  Further work is required to show whether there is a maximum radius for disc fragmentation, which will depend on the equilibrium disc structure produced by magnetised self-gravitating discs with realistic thermodynamics.

Our increased fragment masses are expected from analytic calculations of the magnetic Jeans mass \citep{Strittmatter1966}.  For a medium with a uniform field along a single axis, a density perturbation will only be resisted along the two axes perpendicular to the field vector.  The Jeans length $\lambda_J$ along these axes is modified thus \citep{Chandrasekhar1953}:

\begin{equation}
\lambda^2_{J,mag} = \lambda^2_J \left(1+ \frac{v^2_A}{c_s^2}\right) = \lambda^2_J \left(1+ \frac{1}{\beta_p}\right)
\end{equation} 

\noindent However, a perturbation of Jeans mass $M_J \propto \rm{\rho} \lambda_J \lambda^2_{J,mag}$ will be insufficient to contract along the magnetic field lines - indeed, a Jeans mass $M_J \propto \rm{\rho} \lambda^3_{J,mag}$ is needed.  This back-of-the-envelope calculation suggests we should expect an increase in fragment mass by a factor of $\left(1+ \frac{1}{\beta_p}\right)^{3/2}$ - between $1.1^{3/2}$ to $2\sqrt{2}$.  We do indeed see that our maximum fragment mass increases by a factor of 2.  Applying the converse calculation suggests that $\beta_p\approx1.7$ in the most massive fragments (when the disc was initialised with $\beta_p=10$).  

A more accurate estimate of the fragment mass should consider the properties of the disc material inside a spiral density wave \citep{Forgan2011a}.  Future studies of expected initial fragment mass should consider the impact of magnetic fields in this manner.

We should be careful about extrapolating fragment behaviour from numerical experiments like this - the fragments are interacting with a disc structure unlikely to exist in real discs.  In practice, we expect $\beta_c$ to increase with increasing proximity to the star, as the cooling time of the gas increases with increasing $\Sigma$.  This being said, we do see rapid inward migration, which is a common feature of fragments in self-gravitating discs, not only due to the increased torques exerted on the fragment from a more massive disc \citep{Baruteau2011}, but also the fragment's inability to open a gap and enter the slower Type II migration mode \citep{Malik2015a}.  We recommend CLUMPFIND analysis techniques be applied to simulations of fragmenting discs with realistic thermodynamics to investigate the dynamical properties of fragments further (see e.g. Hall et al, in prep).

Of particular importance to fragment survival is the evolution of spin angular momentum.  \citet{Kim2003}, in their study of giant molecular cloud formation in magnetised galactic discs, noted that the GMCs formed in the disc were subject to strong magnetic braking, losing a significant portion of their initial spin angular momentum.  In fragments, this will have important consequences for the discs they may host \citep{Forgan2016b}.  It will also affect the fragment's ability to collapse into bound objects, preventing Roche lobe overflow and tidal disruption \citep[cf][]{Boley2010,Nayakshin2010, Zhu2012,TD_synthesis}.

We must note that none of our simulations resolve the magneto-rotational instability (MRI).  In protostellar discs, this instability primarily acts in the inner disc, most especially in regions where artificial viscosity dominates our simulations.  In the case of a global SPH disc simulation, we find that resolving the MRI is equivalent to another vertical resolution criterion, which in practice is computationally prohibitive.  For the fastest growing MRI mode to be resolved, we must satisfy:

\begin{equation}
    \frac{h}{H} < 2 \pi \sqrt{\frac{16}{15\beta_p}}
\end{equation}

We do not satisfy this condition anywhere in our discs, even for relatively large $\beta_p$.  As a result, we only resolve a fraction of the MRI modes present, and our description of the magnetic turbulence is incomplete.  We therefore cannot comment on the possible interaction between MRI and GI, in particular whether GI modes can encourage the activation of MRI in the inner disc, which semi-analytic models predict sustains a limit cycle of episodic accretion \citep{Armitage_et_al_01,Martin2014}.  Conversely, simulations have already indicated that MRI modes can weaken the gravitational instability and prevent collapse \citep{Fromang2005,Riols2016}, in much the same way that fragmentation has been suppressed in our simulations.  The fact that we find qualitative agreement with studies that fully resolve the MRI suggests that our results regarding disc fragmentation are robust.

Finally, we should acknowledge the critical role that non-ideal MHD effects may play in self-gravitating discs, especially in the coupling of gas and dust grains.  Spiral density waves in unmagnetised discs have been shown to collect centimetre-metre size grains, in some cases showing significant overdensities \citep{Rice2004,Dipierro2015}.  Whether such overdensities should be expected in observations of self-gravitating discs depends on the ability of grains to grow to sizes such that the Stokes number is of order unity.  This in turn requires the velocity dispersion of the grains to remain sufficiently low during growth, which tends to restrict this effect to large radii \citep{Booth2016}.  This is where magnetic fields are most efficient at reducing gravitational stresses and spiral perturbation amplitudes, so magnetic fields may also suppress grain growth in self-gravitating discs.

Depending on the grain charge, non-ideal effects are likely to play an important role in mediating the separation of the gas and dust phases.  In particular, Ohmic heating is likely to alter the structure of the spiral density wave and the weak shocks it induces.

\section{Conclusions} \label{sec:conclusions}

\noindent We have conducted a series of numerical experiments on magnetised self-gravitating discs using the smoothed particle magnetohydrodynamics code \textsc{phantom}, to test the effects of magnetic fields on disc fragmentation.  We parametrise the disc thermodynamics by fixing the cooling time relative to the dynamical time ($\beta_c$), and we initially impose a purely toroidal magnetic field, where the field strength is set as a constant plasma parameter $\beta_p$, i.e. the magnetic pressure is initially a fixed fraction of the thermal pressure.

We select sufficiently small values of $\beta_c$ that fragmentation occurs for all our discs in the absence of magnetic fields.  We find that fragmentation in magnetised protostellar discs can still occur provided $\beta_c<5$, for both weak and strong fields (i.e. $\beta_p=1$).  For $\beta_c=5$, fragmentation can be suppressed even by relatively weak fields ($\beta_p=10$).

Discs that fragment in the presence of magnetic fields produce fragments of increased mass compared to unmagnetised discs.  Their orbital eccentricity and inclination are very similar to unmagnetised fragments, although our simulations show that there may be a minimum and maximum radius at which fragmentation occurs, as opposed to merely a minimum for unmagnetised discs. 

We recommend that future studies of disc fragmentation must consider even weak magnetic fields, as they are likely to push the fragmentation zone even further from the central object, with obvious implications for planet/brown dwarf formation in protostellar discs, and star formation in discs around active galactic nuclei.

\section*{Acknowledgments}

DHF and IAB gratefully acknowledge support from the ECOGAL project, grant agreement 291227, funded by the European Research Council under ERC-2011-ADG.  DJP gratefully acknowledges funding via grants DP130102078 and FT130100034 and via Future Fellowship FT130100034 from the Australian Research Council.  This work relied on the compute resources of the St Andrews MHD Cluster.  Surface density plots were created using \textsc{SPLASH} \citep{SPLASH}.

\bibliographystyle{mnras} 
\bibliography{magGIdisc}

\label{lastpage}

\end{document}